\documentclass[twocolumn,aps,showpacs,floatfix,superscriptaddress]{revtex4}
\pdfoutput=1 
\usepackage{amsmath,amssymb,eucal,graphicx}
 
\begin{document}
\title{Dynamics of Repulsion Processes}
\author{P. L. Krapivsky}
\affiliation{Physics Department, Boston University, Boston MA 02215, USA}


\begin{abstract}
We study dynamical behaviors of one-dimensional stochastic lattice gases with repulsive interactions whose span can be arbitrary large. We endow the system with a zero-temperature dynamics, so that the hops to the empty sites which would have led to the increase of energy are forbidden. We assume that the strength of interactions sufficiently quickly decreases with the separation between the particles, so that interactions can be treated in a lexicographic order. For such repulsion processes with symmetric nearest-neighbor hopping we analytically determine the density-dependent diffusion coefficient. We also compute the variance of the displacement of a tagged particle. 
\end{abstract}

\maketitle

\section{Introduction}

Lattice models play a central role in equilibrium and non-equilibrium statistical mechanics \cite{Baxter82,Bell89,Spohn91,KL99,D98,S00,BE07,D07,book}. Many lattice models admit the lattice gas representation in which each lattice site is either occupied by a particle or empty, and particles in nearby sites interact according to some specified laws. In equilibrium statistical mechanics, the goal is to compute a free energy as a function of density and temperature, and to understand the properties of phase transitions which often occur in lattice gases in $d\geq 2$ dimensions. In non-equilibrium setting, namely for lattice gases supplemented with stochastic dynamics, interesting dynamical behaviors and rich non-equilibrium steady states arise already in one dimension \cite{Spohn91,KL99,D98,S00,BE07,D07,book}. 

Here we consider one-dimensional lattice gases which are endowed with stochastic dynamics. We analyze various stochastic zero-temperature dynamics which forbid energy-raising moves. Many our results are  exact in the finite setting, viz. for $N$ particles on a ring with $L$ sites, but we focus on the thermodynamic limit, $N\to\infty$ and $L\to\infty$ with $\rho = N/L$ kept finite, which is particularly important. We consider lattice gases with exclusion property (each site contains at most one particle), so that the density lies in the range $0<\rho<1$.  


Among lattice gases, a simple exclusion process (SEP) is the most tractable. In the SEP, 
particles undergo a nearest-neighbor hopping and thereby interact only through the exclusion property (no more than one particle per site). A rich set of behaviors has been discovered for the SEP, see \cite{D98,S00,BE07,D07} and references therein. Here we investigate a repulsion process (RP) where in addition to exclusion, there is an energy cost when particles occupy adjacent sites. We postulate the following dynamics:
\begin{enumerate}
\item If an attempted hop (to a neighboring empty site) would not increase the number of nearest-neighbor pairs of particles, it is always performed. 
\item If a hop would increase of the number of nearest-neighbor pairs of particles, it is never performed.
\end{enumerate}
One can associate a Hamiltonian
\begin{equation}
\label{Ham}
\mathcal{H} = \sum_{i=1}^L n_i n_{i+1}
\end{equation}
with the repulsion process. Here $n_i=1$ if site $i$ is occupied, and $n_i=0$ if it is empty. The above dynamics is based on the exchange, $n_i \leftrightarrow n_{i+1}$. This exchange cannot occur if it would lead to the increase of the number of nearest-neighbor pairs of particles, and thereby to the increase of energy. At zero temperature, energy raising moves are forbidden, so the dynamics underlying the repulsion process is a zero-temperature dynamics associated with Hamiltonian \eqref{Ham}.

The RP is a very basic lattice gas model and it naturally arises in various contexts. For instance, the KLS model proposed by Katz, Lebowitz, and Spohn \cite{KLS84} turns into the RP at a certain point of the parameter space. The KLS model has played a major role in elucidating key conceptual issues in non-equilibrium dynamics \cite{Krug91,Krug01,Zia,Bunin13}. The one-parameter version of the KLS model can be interpreted as the repulsion process with finite-temperature dynamics, and it reduces to the RP at zero temperature and to the SEP at infinite temperature. A number of our results for the RP admit a generalization to the KLS model \cite{Krug01,PK13}, but the tools are very different and one can go much further for the simpler RP model. 

The PR also arises in the context of the Ising model on the square lattice supplemented with zero-temperature spin-flip dynamics  \cite{KO13}. (This was the original motivation for this study.) In a study of the evolution of a quadrant of minority phase, one can map the interface dynamics onto a one-dimensional lattice gas. In the simplest case of nearest-neighbor interactions, the lattice gas is merely the SEP (with symmetric hopping), adding next-nearest-neighbor interactions leads to the RP, adding interactions with 3$^\text{rd}$ nearest neighbors leads to a repulsion process with next-nearest-neighbors, viz. with zero-temperature dynamics associated with Hamiltonian  
\begin{equation}
\label{Ham:2}
\mathcal{H}_2 = J_1\sum_{i=1}^L n_i n_{i+1} + J_2\sum_{i=1}^L n_i n_{i+2} 
\end{equation}

In this paper, we first study the RP with Hamiltonian \eqref{Ham}. We then show how to extend the results to the RP with Hamiltonian \eqref{Ham:2} with arbitrary $J_1>J_2>0$. Finally, we outline the extension to an infinite family of lattice gases class, namely to a class of generalized RPs with interactions of arbitrary finite range. 

The rest of this paper is organized as follows. In the following section we present some of our main results without derivations. In Sect.~\ref{SS}, we determine the structure of steady states and compute correlation functions for the RP with Hamiltonian \eqref{Ham}. In Sect.~\ref{CF}, we compute various correlation functions. The results of Sects.~\ref{SS}--\ref{CF} are then applied to the determination of the diffusion coefficient and to probing the self-diffusion process  (Sect.~\ref{DSD}). In Sect.~\ref{NNN}, we investigate the RP with Hamiltonian \eqref{Ham:2}. The influence of repulsive interactions with longer span is analyzed in Sect.~\ref{GRP}. We conclude with a discussion (Sect.~\ref{Discussion}). Some details of the calculations are presented in Appendices. 

\section{Main Results}
\label{main}

We are chiefly interested in the RP with symmetric hopping. Our methods equally apply to the driven case (when the hopping is biased) since for the RP on the ring the structure of the steady states is independent on the bias and its strength. Hence whenever we present results for the driven RP, we limit ourselves to the totally asymmetric RP in which only hopping to the right are allowed (the rate of allowed hops is set to unity as in the symmetric version). 

The structure of steady states is remarkably simple: They are characterized by the maximal number of islands. When $\rho<1/2$, each occupied site is delimited at both ends by vacant sites. Thus after a transient period the system will be in a configuration like 
\begin{equation}
\label{example}
\bullet\,\circ\,\circ\,\bullet\,\circ\,\bullet\,\circ\,\bullet\,\circ\,\bullet\,\circ\,\circ\,\circ\,\bullet\,\circ\,\bullet\,\circ\,\bullet\,\circ\,\circ\,\bullet\,\circ
\end{equation}
Hereinafter $\bullet$ denotes a particle and $\circ$ denotes a vacancy. (In \eqref{example} we have illustrated the maximal-island configuration on the ring with $N=9$ particles and $V=13$ vacancies.) The system will then forever wander on the phase space of such maximal-island configurations. This phase is clearly connected as follows e.g. by noting that each maximal-island configuration can evolve into the maximal-island configuration containing the longest possible string of alternating particles and vacancies, namely $\bullet\,\circ\,\bullet\,\circ\,\cdots \bullet\,\circ$, followed by the string of vacancies. 

Similarly in the high-density region, $\rho>1/2$, in every steady state configuration each vacant site is delimited at both ends by occupied sites. Low- and high-density regions are treated in the same way and the results are related via the mirror symmetry, $\rho\leftrightarrow 1-\rho$. Hence it suffices to analyze one of the two regions. In the following, we always consider the low-density region if not stated otherwise. 

The remarkable feature of maximal-island configurations is that they all occur with equal probabilities.  Once one establishes these assertions about the structure of steady states configurations and their probabilities, the task of computing the average current, the correlation functions, etc. essentially reduces to a rather straightforward combinatorial analysis. The emerging results are exact for finite systems, so the thermodynamic limit is taken at the end and there are no preliminary assumptions which are often (tacitly) implied in the analysis of infinite-particle systems. 

Here are some of our findings. The probability of each maximal-island configuration is $\mathcal{C}^{-1}$, where $\mathcal{C}$ is the total number of maximal-island configurations with $N$ particles and $V$ vacancies that can be arranged on a ring of size $L=N+V$:
\begin{equation}
\label{conf}
\mathcal{C} = \binom{V}{N} + \binom{V-1}{N-1} = \frac{L}{N}\,\binom{V-1}{N-1}
\end{equation}
in the low-density region $N\leq V$. For the totally asymmetric RP, the steady state current $J$ is
\begin{equation}
\label{current:exact}
J = \frac{\binom{V-2}{N-1}}{\binom{V}{N} + \binom{V-1}{N-1}} = \frac{N}{L}\,\frac{V-N}{V-1}
\end{equation}
In the thermodynamic limit 
\begin{equation}
\label{current:RP}
J(\rho) = 
\begin{cases}
\frac{\rho(1-2\rho)}{1-\rho}  & 0<\rho<\frac{1}{2}\\
\frac{(1-\rho)(2\rho-1)}{\rho}  &\frac{1}{2}<\rho<1
\end{cases}
\end{equation}
where the mirror symmetry allows to deduce the current in the high-density region. 

The pair correlation function is given by 
\begin{equation}
\label{nn:exact}
\langle n_i n_{i+\ell}\rangle = \sum_{q=1}^{\lfloor \ell/2\rfloor} \binom{\ell - q-1}{q-1}\,
 \frac{\binom{V-\ell+q-1}{N-q-1}}{\binom{V}{N} + \binom{V-1}{N-1}} 
\end{equation}
In the thermodynamic limit the pair correlation function, more precisely the connected pair correlation function, has a strikingly simple form
\begin{equation}
\label{nn:simple}
\langle n_i n_j\rangle_c\equiv \langle n_i n_j\rangle -\rho^2 
= \rho(1-\rho) \left(-\frac{\rho}{1-\rho}\right)^{|j-i|}
\end{equation}
valid for all $i$ and $j$. Similarly 
\begin{equation}
\label{nn2:high}
\langle n_i n_j\rangle_c =  \rho(1-\rho) \left(-\frac{1-\rho}{\rho}\right)^{|j-i|}
\end{equation}
in the high-density region, $\rho>\tfrac{1}{2}$. Thus the connected pair correlation function exhibits a pure exponential decay modulated by an oscillating sign. 

We have also computed higher-order correlation functions. They exhibit remarkable factorization properties, e.g. in the low-density regime the three particle correlation function reads
\begin{equation}
\label{nnn:low}
\langle n_i n_j n_k\rangle = \frac{\langle n_i n_j\rangle \langle n_j n_k\rangle}{\langle n_j\rangle}
\end{equation}
for all $i\leq j\leq k$. 

Lattice gases with symmetric hopping usually exhibit a universal {\em hydrodynamic} behavior, namely  on large spatial and temporal scales they are governed by a non-linear diffusion equation \cite{Spohn91}
\begin{equation}
\label{rho:eq}
\frac{\partial \rho}{\partial t} = \frac{\partial }{\partial x}\!\left[D(\rho)\, \frac{\partial \rho}{\partial x}\right]
\end{equation}
The diffusion coefficient $D(\rho)$ represents the spread of disturbances on an equilibrium background with uniform density $\rho$. The diffusion coefficient {\em depends} on the density in most lattice gases thereby making \eqref{rho:eq} a non-linear partial differential equation. Thus all the microscopic details of lattice gas dynamics are absorbed into a single number, the diffusion coefficient $D(\rho)$. 

The diffusion description has been justified for numerous lattice gas models \cite{Spohn91,KL99}, but the diffusion coefficient $D(\rho)$ has been computed in a very few models, mostly in the situations when the diffusion coefficient is independent on the density. We have found that for the RP
\begin{equation}
\label{diff:RP}
D(\rho) = 
\begin{cases}
(1-\rho)^{-2}    & 0<\rho<\frac{1}{2}\\
\rho^{-2}          &\frac{1}{2}<\rho<1
\end{cases}
\end{equation}

The limiting behaviors, $D(\rho=0)=D(\rho=1)=1$ in \eqref{diff:RP}, can be understood without calculations. When $\rho\to 0$, it suffices to analyze a single particle in the empty lattice. This particle diffuses with $D=1$ (recall that we have set the hopping rates to unity). Similarly when the density is very close to the maximal density $\rho=1$, we can consider a single vacancy in otherwise fully packed lattice. To establish $D(\tfrac{1}{2})=4$, let us slightly perturb the $\rho=\frac{1}{2}$ state. There is just one half-filled equilibrium configuration, $\ldots\bullet\circ\bullet\circ\bullet\circ\bullet\circ\ldots$. Consider its smallest perturbation  
\begin{equation*}
\ldots\bullet\circ\bullet\circ\bullet\circ\bullet\bullet\circ\bullet\circ\bullet\circ\bullet\circ\ldots
\end{equation*}
with a single doublet. In this configuration both particles forming the doublet can hop. A single hopping event, say to the right, leads to 
\begin{equation*}
\circ\bullet\bullet\circ\bullet\circ ~\Longrightarrow~ \circ\bullet\circ\bullet\bullet\circ
\end{equation*}
Therefore the doublet effectively undergoes a symmetric hopping on distance two, so its diffusion coefficient is 4 times larger than the bare diffusion coefficient of a particle. 

Apart from diffusion, we analyze self-diffusion: We tag a particle and probe its long time behavior. For the one-dimensional RP (and other lattice gases with nearest-neighbor hopping and exclusion), the mean-square displacement of the tagged particle grows as 
\begin{equation}
\label{SD_1d}
\langle X^2(t)\rangle = \mathcal{D}(\rho)\,\sqrt{t}
\end{equation}
We have found that the amplitude $\mathcal{D}(\rho)$ is given by 
\begin{equation}
\label{self-diff:RP}
\mathcal{D}(\rho) = \frac{2}{\sqrt{\pi}}\,\frac{1}{\rho^2}\times
\begin{cases}
\rho(1-2\rho)                & 0<\rho<\frac{1}{2}\\
(1-\rho)(2\rho-1)          &\frac{1}{2}<\rho<1
\end{cases}
\end{equation}
for the RP. In contrast to the diffusion coefficient which is maximal at half-filling, the self-diffusion amplitude vanishes at half-filling. 

We now turn to the generalized repulsion processes (GRPs) which are characterized by the Hamiltonian
\begin{equation}
\label{Ham:GRP}
\mathcal{H}_m = J_1\sum_{i=1}^L n_i n_{i+1} +\ldots
+ J_m\sum_{i=1}^L n_i n_{i+m}
\end{equation}
The coupling constants $J_1, \ldots, J_m$ are assumed to be positive, so that interactions between particles separated by distance $\leq m$ are all repulsive; particles separated by larger distance do not interact. In addition we assume that the coupling constants obey the set of constraints
\begin{equation}
\label{JJJ}
J_k>J_{k+1}+\ldots+J_m, \quad k=1,\ldots, m -1
\end{equation}
An attempted hop to the neighboring empty site is again performed if the energy is not raised; otherwise, it is rejected.  The constraints \eqref{JJJ} simplify the analysis since they allow us to treat interactions in a lexicographic order. Namely, if the attempted hop would lead to the decrease of the number of nearest-neighbor pairs, it is always performed; if the number would increase, it is never perform. If the number would remain the same, one should check how the number of next-nearest-neighbors would be affected. If the attempted hop would lead to the increase or decrease of this number, the fate of the hop is settled (it is rejected or accepted, respectively); otherwise we go to the third level. 

There are $m$ relevant low-density regions with different behaviors 
\begin{equation}
\label{regions}
0<\rho<\tfrac{1}{m+ 1}\,, \quad \tfrac{1}{m + 1}<\rho<\tfrac{1}{m}\,, ~~\ldots, 
~~\tfrac{1}{3}<\rho<\tfrac{1}{2}
\end{equation}
In each region, a certain class of maximal-island configurations constitutes the phase space of steady states; in each class, the maximal-island configurations occur with equal probabilities. The corresponding maximal-island configurations are specified by the allowed range for the length $s$ of islands of vacant sites:
\begin{equation}
\label{islands}
\begin{split}
0<\rho<\tfrac{1}{m + 1}                  \qquad   &s\geq m\\
\tfrac{1}{m+ 1}<\rho<\tfrac{1}{m}   \qquad  &s=(m -1, m)\\
\tfrac{1}{m}<\rho < \tfrac{1}{m -1} \qquad  &s=(m -2, m-1)
\end{split}
\end{equation}
etc. ending with region $\tfrac{1}{3}<\rho<\tfrac{1}{2}$  where $s=(1,2)$. Regions $0<\rho<\tfrac{1}{m + 1}$ and $\tfrac{1}{m+ 1}<\rho<\tfrac{1}{m}$ are new, other regions already appear in the analysis of the model with repulsion range $m-1$. Thus in the $\tfrac{1}{k}<\rho<\tfrac{1}{2}$ density range, the behavior stabilizes when $m\geq k$. 

The total number of admissible maximal-island configurations is a generalization of \eqref{conf} in the region with smallest density
\begin{equation}
\label{conf:tiny}
\mathcal{C} = \frac{L}{N}\, \binom{V-(m -1)N-1}{N-1}, \qquad 0<\rho<\tfrac{1}{m + 1}
\end{equation}
In the remaining low-density regions 
\begin{equation}
\label{conf:small}
\mathcal{C} = \frac{L}{N}\, \binom{N}{V-(k -1)N}, \qquad \frac{1}{k+1}<\rho<\frac{1}{k}
\end{equation}
where $k=2,3,\ldots, m$. 

Most of previous results for the RP (which corresponds to $m=1$) admit an extension to the GRPs. Formulas for the single low-temperature region must be replaced by their analogs for the $m$ low-temperature regions \eqref{regions}, but extensions tend to be straightforward. For instance, for the totally asymmetric GRP, the steady state current in the low-density regions is given by
\begin{equation}
\label{current:GRP}
J(\rho) =
\begin{cases}
\frac{\rho[1-(m + 1)\rho]}{1- m \rho}    &  0<\rho<\frac{1}{m + 1}\\
\frac{[(k + 1)\rho-1][1-k\rho]}{\rho}       &  \frac{1}{k+1}<\rho<\frac{1}{k}
\end{cases}
\end{equation}
where $k=2,3,\ldots, m$. 

The behavior of the diffusion coefficient characterizing the GRP with symmetric hopping is even simpler, for every $m\geq 2$ there are just 4 different regions:
\begin{equation}
\label{diff:GRP}
D(\rho) = 
\begin{cases}
(1-m\rho)^{-2}        & 0<\rho<\frac{1}{m+ 1}\\
\rho^{-2}                 & \frac{1}{m+ 1}<\rho<\frac{1}{2}\\
(1-\rho)^{-2}           & \frac{1}{2}<\rho<\frac{m}{m+1}\\
(m\rho-m+1)^{-2}   &\frac{m}{m+1}<\rho<1
\end{cases}
\end{equation}
The self-diffusion amplitude in the low-density region is 
\begin{equation*}
\label{self-diff:GRP}
\rho^2\mathcal{D} = \frac{2}{\sqrt{\pi}}\times
\begin{cases}
\rho[1-(m+1)\rho]                   & 0<\rho<\frac{1}{m+1}\\
[(k + 1)\rho-1][1-k\rho]           &  \frac{1}{k+1}<\rho<\frac{1}{k}
\end{cases}
\end{equation*}
where again $k=2,3,\ldots, m$. The behavior in the high-density regime is obtained by replacing $\rho\to 1-\rho$ on the right-hand side. For instance,
\begin{equation*}
\mathcal{D} = \frac{2}{\sqrt{\pi}}\,\frac{[k-(k + 1)\rho][k\rho-(k-1)]}{\rho^2}
\end{equation*}
when $\tfrac{k-1}{k}<\rho<\tfrac{k}{k+1}$, where $k=2,3,\ldots, m$. 

The results are particularly neat for the GRPs with infinite interaction range. The constraints \eqref{JJJ} are obeyed when e.g. the coupling constants decay exponentially with distance, $J_k=\delta^k$ with $\delta\leq \frac{1}{2}$. For such GRPs with totally asymmetric hopping, the current is given by
\begin{equation}
\label{current:GRP_inf}
J(\rho) =
\begin{cases}
\frac{[(k + 1)\rho-1][1-k\rho]}{\rho}               &  \frac{1}{k+1}<\rho<\frac{1}{k}\\
\frac{[k-(k + 1)\rho][k\rho-(k-1)]}{1-\rho}       &  \frac{k-1}{k}<\rho<\frac{k}{k+1}
\end{cases}
\end{equation}
with $k=2, 3, \ldots$. For the same model with symmetric hopping, the diffusion coefficient is
\begin{equation}
\label{diff:GRP_inf}
D(\rho) = 
\begin{cases}
\rho^{-2}                 & 0<\rho<\frac{1}{2}\\
(1-\rho)^{-2}           & \frac{1}{2}<\rho<1
\end{cases}
\end{equation}
while the self-diffusion amplitude is given by
\begin{equation}
\label{self-diff:GRP_inf}
\mathcal{D} = \frac{2}{\sqrt{\pi}}
\begin{cases}
\frac{[(k + 1)\rho-1][1-k\rho]}{\rho^2}          &  \frac{1}{k+1}<\rho<\frac{1}{k}\\
\frac{[k-(k + 1)\rho][k\rho-(k-1)]}{\rho^2}       &  \frac{k-1}{k}<\rho<\frac{k}{k+1}
\end{cases}
\end{equation}
with $k=2, 3, \ldots$. Comparing \eqref{diff:GRP_inf} with \eqref{diff:RP} we see a surprising formal duality  between the density-dependence of the diffusion coefficient for the RP with infinite-range interaction range and for the simplest RP with nearest-neighbor interactions.

\section{Steady States}
\label{SS}

In this section we explain the structure of steady states outlined in Sect.~\ref{main}, and we derive the announced results for the number of maximal-island configurations, Eq.~\eqref{conf}, and for the steady state current in the totally asymmetric case, Eq.~\eqref{current:exact}. 

We put our stochastic lattice gas on the ring and let it evolve. The initial condition will be eventually forgotten. To appreciate the emergence of configurations like \eqref{example} we notice that in each hopping event the number of islands either increases or remains the same; the decrease of the number of islands would imply the increase of the energy which is forbidden. Thus the system eventually reaches a state where the number of islands is maximal and then it wanders on the phase space of such maximal-island configurations. To prove that all maximal-island configurations are equally probable we denote by $P(C)$ the probability of being in maximal-island configuration $C$ and write the stationarity condition 
\begin{equation}
\label{stationary}
P(C)\sum_{C'}T(C\rightarrow C')=\sum_{C''}P(C'')T(C''\rightarrow C)
\end{equation}
where $T(C\rightarrow C')$ is the transfer rate from $C$ to $C'$. Since $T=1$ if the evolution is allowed and $0$ otherwise, we need to count the number of ways into and out of a configuration.  Consider for concreteness the totally asymmetric version of the RP. The basic hopping event $\circ\bullet\circ\circ\rightarrow\circ\circ\bullet\circ$ shows that the evolution out of $C$ occurs at the left edges of islands of vacancies of length $\geq 2$. The same basic process shows that the evolution into $C$ is counted by the number of right edges of islands of vacancies of length $\geq 2$. Therefore $\sum_{C'}T(C\rightarrow C')=\sum_{C''}T(C''\rightarrow C)$ implying that if $P(C)$ are equal for all configurations, Eq.~\eqref{stationary} is satisfied. 

To determine $\mathcal{C}$ we proceed in the same way as in \cite{GKR}. We pick up an arbitrary site as a first site and represent configurations as strings. Configurations in which the first site is empty look like 
\begin{equation*}
\circ\,\,\overbrace{\downarrow\underbrace{\circ\downarrow\circ\downarrow\circ\downarrow\circ\downarrow\circ\downarrow\circ\downarrow\circ\downarrow\circ}_{V-1 ~\text{vacancies}}\downarrow}^{N~ \text{particles}}
\end{equation*}
with $N$ particles distributed among $V$ possible positions (each such position is denoted by $\downarrow$). The number of such configurations is $\binom{V}{N}$. Similarly configurations with occupied first site look like
\begin{equation*}
\bullet\,\circ\,\,\overbrace{\downarrow\underbrace{\circ\downarrow\circ\downarrow\circ\downarrow\circ\downarrow\circ\downarrow\circ\downarrow\circ}_{V-2 ~\text{vacancies}}\downarrow}^{N-1~ \text{particles}}\,\,\circ
\end{equation*}
and the number of such configurations is $\binom{V-1}{N-1}$. Combining these two contributions we arrive at \eqref{conf}.

To compute the steady state current $J$ we count the number of configurations of the type
\begin{equation*}
\circ\,\bullet\,\circ\,\circ\,\,\overbrace{\downarrow\underbrace{\circ\downarrow\circ\downarrow\circ\downarrow\circ\downarrow\circ\downarrow\circ}_{V-3 ~\text{vacancies}}\downarrow}^{N-1~ \text{particles}}
\end{equation*}
and divide it by the total number of maximal-island configurations given by Eq.~\eqref{conf}. This leads to  Eq.~\eqref{current:exact}.

\section{Correlation functions}
\label{CF}

In this section we compute correlation functions. The results for the pair correlation functions, Eqs.~\eqref{nn:exact}--\eqref{nn2:high}, have been announced in Sect.~\ref{main}. We also establish \eqref{nnn:low} and show how to compute other higher-order correlation functions. 

Let us first compute the pair correlation function $\langle n_i n_j\rangle$ in the low-density regime. In the steady state, the pair correlation depends only on the separation $\ell=j-i$ between the sites. Obviously
\begin{equation}
\label{nn1} 
\langle n_i^2\rangle=\langle n_i\rangle=\rho, \qquad \langle n_i n_{i+1}\rangle=0
\end{equation}
The pair correlation function becomes non-trivial when $\ell=2$. This describes two particles separated by one site (which is necessarily empty in the low-density regime). We need to count the maximal-island configurations of the type
\begin{equation}
\label{L2:conf} 
\circ\bullet\circ\bullet\circ\overbrace{\downarrow\underbrace{\circ\downarrow\circ\downarrow\circ\downarrow\circ\downarrow\circ\downarrow\circ}_{V-3 ~\text{vacancies}}\downarrow}^{N-2~ \text{particles}}
\end{equation}
One gets $\binom{V-2}{N-2}$, from which
\begin{equation}
\label{nn2} 
\langle n_i n_{i+2}\rangle = \frac{\binom{V-2}{N-2}}{\binom{V}{N} + \binom{V-1}{N-1}} 
\to \frac{\rho^2}{1-\rho}
\end{equation}

Generally to compute $\langle n_i n_{i+\ell}\rangle$ let us consider configurations with $q-1$ particles and $\ell - q$ vacancies between sites $i$ and $i+\ell$:
\begin{equation}
\label{LL:conf} 
\circ\bullet\,\underbrace{\circ\overbrace{\downarrow\circ\downarrow\circ\downarrow\circ\downarrow}^{q-1~ \text{particles}}\circ}_{\ell -q~\text{vacancies}}\, \bullet\circ\,\overbrace{\downarrow\underbrace{\circ\downarrow\circ\downarrow\circ\downarrow\circ\downarrow\circ\downarrow\circ}_{V-\ell+q-2 ~\text{vacancies}}\downarrow}^{N-q-1~ \text{particles}}
\end{equation}
Particles at sites $i$ and $i+\ell$ are explicitly presented in \eqref{LL:conf}, possible locations of other particles are shown by $\downarrow$. The number of configurations of type \eqref{LL:conf} is
\begin{equation}
\label{conf:q}
\binom{\ell - q-1}{q-1}\binom{V-\ell+q-1}{N-q-1}
\end{equation}
The first binomial factor accounts for placing $q-1$ particles into $\ell - q-1$ possible positions inside the string $(i, i+1,\ldots,i+\ell)$. The second binomial factor describes all possible placings of $N-q-1$ particles into $V-\ell+q-1$ admissible positions outside the string $(i, i+1,\ldots,i+\ell)$. Taking into account that $q$ varies from $q=1$ to $q=\lfloor \tfrac{\ell}{2}\rfloor$ we arrive at \eqref{nn:exact}. In the thermodynamic limit, Eq.~\eqref{nn:exact} becomes
\begin{equation}
\label{2_L}
\langle n_i n_{i+\ell}\rangle = \sum_{q=1}^{\lfloor \ell/2\rfloor} \binom{\ell - q-1}{q-1}\,
\frac{\rho^{q+1} (1-2\rho )^{\ell-2q}}{(1-\rho)^{\ell-q}}
\end{equation}
In particular, for $3\leq \ell\leq 7$ the pair correlation function reads
\begin{equation}
\label{2CF}
\begin{split}
\frac{\langle n_i n_{i+3}\rangle}{\rho^2}\! &= \frac{1-2\rho}{(1-\rho)^2}\\
\frac{\langle n_i n_{i+4}\rangle}{\rho^2}\! &= \frac{(1-2\rho)^2}{(1-\rho)^3}+ \frac{\rho}{(1-\rho)^2}\\
\frac{\langle n_i n_{i+5}\rangle}{\rho^2}\! &= \frac{(1-2\rho)^3}{(1-\rho)^4}+ \frac{2\rho(1-2\rho)}{(1-\rho)^3}\\
\frac{\langle n_i n_{i+6}\rangle}{\rho^2}\! &= 
\frac{(1-2\rho)^4}{(1-\rho)^5}+ \frac{3\rho(1-2\rho)^2}{(1-\rho)^4} + \frac{\rho^2}{(1-\rho)^3}\\
\frac{\langle n_i n_{i+7}\rangle}{\rho^2}\! &= 
\frac{(1-2\rho)^5}{(1-\rho)^6}+ \frac{4\rho(1-2\rho)^3}{(1-\rho)^5} + \frac{\rho^2(1-2\rho)}{(1-\rho)^4}
\end{split}
\end{equation}

The remarkable feature of the SEP is that the correlation functions factorize: $\langle n_{i} n_{j}\rangle = \langle n_{i}\rangle \langle n_{j}\rangle=\rho^2$, etc. We have written \eqref{2CF} in a way that emphasizes that for the RP, the pair correlation function does not factorize. Despite this lack of factorization, the connected pair correlation is given by a simple formula \eqref{nn:simple}. This is easily verified for $\ell\leq 7$ using explicit expressions \eqref{nn1}, \eqref{nn2}, and \eqref{2CF}. To derive Eq.~\eqref{nn:simple} from Eq.~\eqref{2_L} in the general case, consider first the situation when the distance $\ell$ between the sites is even. Writing $\ell=2k+2$ and $q=1+p$, we transform the sum on the right-hand side of \eqref{2_L} into
\begin{equation*}
\frac{\rho^2}{1-\rho}\left(\frac{1-2\rho}{1-\rho}\right)^{2k}\sum_{p=0}^k\binom{2k-p}{p}x^p, \quad
x=\frac{\rho(1-\rho)}{(1-2\rho)^2}
\end{equation*}
The last sum can be computed for arbitrary $k$ 
\begin{eqnarray*}
\sum_{p=0}^k\binom{2k-p}{p}x^p &=& \frac{1}{2}\left(1+\frac{1}{\sqrt{4x+1}}\right)X_+^k \\
                                                      &+& \frac{1}{2}\left(1-\frac{1}{\sqrt{4x+1}}\right)X_-^k
\end{eqnarray*}
where
\begin{equation*}
X_\pm = \frac{2x+1\pm \sqrt{4x+1}}{2}
\end{equation*}
Combining all these results we establish Eq.~\eqref{nn:simple} in the situation when the distance $\ell=j-i$ between the sites is even. The situation when the distance between the sites is odd is treated in a similar manner. 

We now compute higher-order correlation functions. We begin with the three particle correlation function. We shall use the shorthand notation $\langle 123\rangle \equiv \langle n_{i_1} n_{i_2} n_{i_3}\rangle$. Let $\ell=i_2-i_1>0$ and $m=i_3-i_2>0$ be the separations between the particles. Configurations 
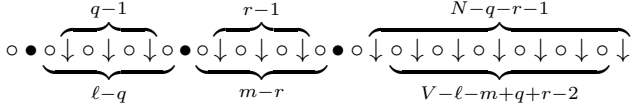
\begin{equation*}
\circ\bullet\underbrace{\circ\overbrace{\downarrow\circ\downarrow\circ\downarrow}^{q-1}\circ}_{\ell -q} \bullet\underbrace{\circ\overbrace{\downarrow\circ\downarrow\circ\downarrow}^{r-1}\circ}_{m -r} \bullet
\circ\overbrace{\downarrow\underbrace{\circ\downarrow\circ\downarrow\circ\downarrow\circ\downarrow\circ\downarrow\circ}_{V-\ell-m+q+r-2}\downarrow}^{N-q-r-1}
\end{equation*}
represent arrangements with $q-1$ particles between sites $i_1$ and $i_2$ and $r-1$ particles between sites $i_2$ and $i_3$. The total number of such configurations is
\begin{equation*}
\binom{\ell - q-1}{q-1} \binom{m - r-1}{r-1}\binom{V-\ell-m+q+r-1}{N-q-r-1}
\end{equation*}
so that
\begin{eqnarray*}
\langle 123\rangle &=& \sum_{q=1}^{\lfloor \ell/2\rfloor}\sum_{r=1}^{\lfloor m/2\rfloor}
\binom{\ell - q-1}{q-1} \binom{m - r-1}{r-1}\\
&\times&  \frac{\binom{V-\ell-m+q+r-1}{N-q-r-1}}{\binom{V}{N} + \binom{V-1}{N-1}} 
\end{eqnarray*}
which in the thermodynamic limit becomes
\begin{eqnarray*}
\frac{\langle 123\rangle}{\rho} &=& 
\sum_{q=1}^{\lfloor \ell/2\rfloor} \binom{\ell - q-1}{q-1}\, \frac{\rho^{q} (1-2\rho )^{\ell-2q}}{(1-\rho)^{\ell-q}}\\
&&\sum_{r=1}^{\lfloor m/2\rfloor}
\binom{m - r-1}{r-1}\, \frac{\rho^{r} (1-2\rho )^{m-2r}}{(1-\rho)^{m-r}}
\end{eqnarray*}
The first sum on the right-hand side of the above equation is equal to $\rho^{-1}\langle 12\rangle$, while the second sum is equal to $\rho^{-1}\langle 23\rangle$. (Here $\langle 12\rangle=\langle n_{i_1}n_{i_2}\rangle$ and $\langle 23\rangle=\langle n_{i_2} n_{i_3}\rangle$; also $\rho=\langle 1\rangle=\langle 2\rangle=\langle 3\rangle$.)
Thus a seemingly cumbersome expression for the three particle correlation function can be re-written as \begin{equation}
\label{3CF}
\langle 123\rangle = \frac{\langle 12\rangle\,\langle 23\rangle}{\langle  2\rangle}
\end{equation}
which is Eq.~\eqref{nnn:low} announced in Sect.~\ref{main}.
Equation \eqref{3CF}  is reminiscent to the Kirkwood's superposition approximation \cite{Kirkwood35}    which is often used in liquid theory \cite{Balescu,RL77}. In the present case, however, it is an equality rather than an uncontrolled approximation.  

For the RP, the correlation functions do not factorize as we already know. However, Eq.~\eqref{3CF} shows that the three-particle correlation functions factorize into the product of pair correlation functions. This remarkable property continues to hold for higher-order correlation functions: The same calculation as above shows that 
\begin{equation}
\label{high_CF}
\left\langle \prod_{a=1}^k n_{i_a}\right\rangle = \frac{1}{\rho^{k-2}}\prod_{a=1}^{k-1}
\left\langle n_{i_a} n_{i_{a+1}}\right\rangle
\end{equation}
We could have written the denominator on the right-hand side of \eqref{high_CF} as $\rho^{k-2}=\prod_{2\leq a\leq k-1}\langle n_{i_a}\rangle$, so that like in \eqref{3CF} the $n_{i_a}$ factors which appear twice in the numerator would be balanced by appearance in the denominator thereby agreeing with a single appearance on the left-hand side of \eqref{high_CF}. 

By inserting \eqref{nn:simple} into \eqref{high_CF} we recast the higher-order correlation functions into a sum of exponential factors. The simple form of Eq.~\eqref{nn:simple} suggests that the connected correlation functions defined via  \cite{Balescu}
\begin{equation}
\langle n_{i_1} \cdots n_{i_a}\rangle_c = \left\langle (n_{i_1}-\rho)\cdots(n_{i_a}-\rho)\right\rangle
\end{equation}
may have a simpler form. The three-particle connected correlation function
\begin{equation*}
\langle 123\rangle_c = \langle 123\rangle 
- \rho\left[\langle 12\rangle + \langle 23\rangle + \langle 13\rangle\right]  +  2\rho^3
\end{equation*}
simplifies to a single exponential 
\begin{equation}
\label{nnn:simple}
\langle 123\rangle_c = \rho(1-\rho)(1-2\rho)\left(-\frac{\rho}{1-\rho}\right)^{i_3-i_1}
\end{equation}
In Eq.~\eqref{nnn:simple} we assume that $i_1\leq i_2\leq i_3$; when some indexes are equal, the predictions remain correct. Similar remarks apply to following formulae with more indexes, e.g., to Eqs.~\eqref{nnnn}, \eqref{n5}, \eqref{n6}.

The four-particle connected correlation function reduces to a combination of two exponential terms
\begin{eqnarray}
\label{nnnn}
\langle 1234 \rangle_c &=&
(1-2\rho)^2 \rho(1-\rho)\left(-\frac{\rho}{1-\rho}\right)^{i_4 - i_1}\nonumber\\
&+& [\rho(1-\rho)]^2 \left(-\frac{\rho}{1-\rho}\right)^{i_4-i_3+i_2-i_1}
\end{eqnarray}

Equations \eqref{nnn:simple}--\eqref{nnnn} indicate that the number of exponential terms increases with order. Explicit formulae for $\langle 12345 \rangle_c$ and $\langle 123456 \rangle_c$ [Eqs.~\eqref{n5} and \eqref{n6} in Appendix~\ref{higher}] confirm this assertion. Generally the number of exponential terms appearing in the $p-$particle connected correlation function is equal to the Fibonacci number $F_{p-1}$ (Appendix~\ref{higher}). Hence the number of exponential factors varies as $g^{p}$, where $g=\tfrac{1}{2}(\sqrt{5}+1)\approx 1.618$ is the golden ratio. In comparison with $2^{p-1}$ exponential factors which appear in ordinary correlation function $\langle 12\ldots p\rangle$, there is a huge cancellation for the connected correlation function $\langle 12\ldots p\rangle_c$, although the growth with $p$ remains exponential.

The behavior in the high-density region is established by replacing $n_i\to 1-n_i$ and $\rho\to 1-\rho$. For instance, the Kirkwood relation \eqref{3CF} becomes
\begin{eqnarray*}
&&\langle (1-n_{i_1})(1- n_{i_2})(1- n_{i_3})\rangle\\
&& = \frac{\langle (1-n_{i_1})(1- n_{i_2})\rangle\,
\langle (1- n_{i_2})(1- n_{i_3})\rangle}{\langle  1-n_{i_2}\rangle}
\end{eqnarray*} 
This and other similar results are a bit more cumbersome than the corresponding expressions in the low-density region when one expresses them in terms of particle correlation functions; in terms of the vacancy correlation functions, however, they are as simple. 

Standard correlation functions do not exploit the nature of the steady states characterizing the RP. For the RP it makes sense to compute the density of islands of vacant sites (in the low-density region). The density $E_s$ of islands of $s$ empty sites delimited at both ends by particles can be written as $\langle n_0(1-n_1)\ldots(1-n_s)n_{s+1}\rangle$ in terms of the standard correlation functions, yet it is easier to compute this quantity directly. Indeed, we need to count the total number of configurations of the type
\begin{equation}
\label{island_s}
\bullet\,\underbrace{\circ\,\circ\,\circ\,\circ\,\circ\,\,\circ}_{s~\text{vacancies}}\,\,\bullet\,\,\underbrace{\circ\,\overbrace{\downarrow\circ\downarrow\circ\downarrow\circ\downarrow\circ\downarrow\circ\downarrow\circ\downarrow}^{N-2~ \text{particles}}\,\circ}
_{V-s ~\text{vacancies}}
\end{equation}
Two particles delimiting the island of $s$ empty sites are explicitly presented in \eqref{island_s}, possible locations of other particles are shown by $\downarrow$. Thus
\begin{equation}
E_s = \frac{N}{L}\,\frac{\binom{V-s-1}{N-2}}{\binom{V-1}{N-1}} = 
\frac{\rho^2}{1-\rho}\left(\frac{1-2\rho}{1-\rho}\right)^{s-1}
\end{equation}
It is also clear that neighboring islands are uncorrelated, e.g. the density $E_{s_1,s_2}$ of the two-island pattern
\begin{equation*}
\bullet\,\underbrace{\circ\,\circ\,\circ\,\circ\,\circ\,\circ\,\circ}_{s_1~\text{vacancies}}\,\bullet\,\underbrace{\circ\,\circ\,\circ\,\circ\,\circ\,\circ\,\circ\,\,\circ}_{s_2~\text{vacancies}}\,\,\bullet\,\,
\end{equation*}
is $E_{s_1,s_2}=\rho^{-1}E_{s_1} E_{s_2}$. The density of $p-$island patterns is similarly 
\begin{equation}
E_{s_1,\ldots,s_p}=\rho^{-(p-1)}\prod_{a=1}^p E_{s_a}
\end{equation}

\section{Diffusion and Self-Diffusion}
\label{DSD}

In this section we establish our chief analytical result \eqref{diff:RP} for the density-dependent diffusion coefficient. We also derive \eqref{self-diff:RP} which gives the amplitude of self-diffusion. We consider the symmetric version: Hopping to the left and right occur with the same unit rate. 

It is generally impossible to compute $D(\rho)$ for interacting lattice gases. The symmetric simple exclusion process (SSEP) is a rare exception: In this case the averages $\langle n_k(t)\rangle$ satisfy a closed equation which on the hydrodynamic scale reduces to Eq.~\eqref{rho:eq} with $D=1$. For interacting lattice gases the governing equations for the averages involve higher-order correlation functions, and this hierarchical nature prevents the derivation of the diffusion coefficient. To circumvent this obstacle, one can try to compute the diffusion coefficient via the Green-Kubo formula \cite{RL77}. This formula expresses the diffusion coefficient through the current-current correlation function. Current-current correlations are still poorly understood for (deterministic) mechanical systems \cite{RL77}. For (stochastic) lattice gases these correlations are more tractable \cite{Spohn91}. The Green-Kubo formula can be schematically written in the form
\begin{equation}
\label{EGK}
D(\rho) = \frac{1}{2\chi(\rho)}\left[H(\rho)-\int_0^\infty dt\,C(t)\right]
\end{equation}
where $H(\rho)$ is the average total hopping rate and 
\begin{equation}
\label{compress:def}
\chi(\rho) = \sum_{\ell=-\infty}^\infty \langle n_0 n_\ell\rangle_c
\end{equation}
is the compressibility (more precisely, the latter is proportional to $\chi$). An explicit expression for $C(t)$ can be found in Ref.~\cite{Spohn91}. This integral contribution has never been computed, apart from a few cases where it has been proven to be equal to zero \cite{Spohn91}. Fortunately, it happens to be the case for the RP (see Appendix~\ref{gradient}), so we can drop this term. 

The average total hopping rate from a site in the symmetric version is two times larger than the current in the asymmetric version. This gives
\begin{equation}
\label{EGK:simple}
D(\rho) = \frac{J(\rho)}{\chi(\rho)}
\end{equation}
which allows us to compute the diffusion coefficient. Using \eqref{nn:simple} and \eqref{nn2:high} we determine the compressibility
\begin{equation}
\label{chi:RP}
\chi(\rho) = \rho(1-\rho) |1-2\rho|  
\end{equation}
Plugging \eqref{current:RP} and \eqref{chi:RP} into \eqref{EGK:simple} we arrive at \eqref{diff:RP}.

Consider now the process of self-diffusion, that is, we tag a particle and follow its trajectory. The lattice gas is assumed to be at equilibrium and the tagged particle is indistinguishable from other particles. Without loss of generality, we can assume that the tagged particle is initially at the origin. Generally for an arbitrary lattice gas with symmetric hopping and in an arbitrary spatial dimension $d$, the average position of the tagged particle does not change with time
\begin{equation}
\label{av_tag}
\langle {\bf X}(t)\rangle = {\bf 0}
\end{equation}
while the mean-square displacement exhibits a diffusive growth
\begin{equation}
\label{var_tag}
\langle {\bf X}^2(t)\rangle = 2d D_s(\rho,d)\,t
\end{equation}
The coefficient of self-diffusion $D_s(\rho,d)$ is unknown even for the SSEP when the spatial dimension is $d\geq 2$. One should also keep in mind that the diffusion of the tagged particle in higher dimensions is generally described by the self-diffusion matrix, so one should replace \eqref{var_tag} by matrix generalization.

The coefficient of self-diffusion may vanish in one dimension. This happens for the SSEP where the mean-square displacement was found \cite{Harris_65,Levitt_73,PMR_77,AP_78,A_83,van_83} to exhibit a remarkable sub-diffusive growth, $\langle X^2\rangle = \mathcal{D}_\text{SSEP}(\rho)\,\sqrt{t}$,  with 
\begin{equation}
\label{var_tag_1d}
\mathcal{D}_\text{SSEP}(\rho)=\frac{2}{\sqrt{\pi}}\,\frac{1-\rho}{\rho}
\end{equation}
As for normal self-diffusion, $P(X,t)=\text{Prob}(X(t)=X)$ is a Gaussian distribution \cite{A_83}. The anomalously slow $\sqrt{t}$ growth is caused by the fact that the original particle ordering is forever preserved for the SSEP in one dimension. The same growth is valid for other exclusion processes with symmetric hopping satisfying two constraints:
\begin{enumerate}
\item No more than one particle per site.
\item Only nearest-neighbor jumps are allowed.
\end{enumerate}

The amplitude $\mathcal{D}(\rho)$ generally depends on the details of the process. The derivation of \eqref{var_tag_1d} given in Refs.~\cite{AP_78,van_83} suggests that the amplitude $\mathcal{D}(\rho)$ can be expressed through the diffusion coefficient $D(\rho)$ and the compressibility $\chi(\rho)$:
\begin{equation}
\label{var_tag_conj}
\mathcal{D}(\rho)= \frac{2}{\sqrt{\pi}}\,\frac{\chi(\rho)}{\rho^2}\,\sqrt{D(\rho)}
\end{equation}
Substituting \eqref{diff:RP} and \eqref{chi:RP} into \eqref{var_tag_conj} we arrive at \eqref{self-diff:RP}. 

Comparing the RP and the SSEP we see that the self-diffusion is faster in the latter model, $\mathcal{D}(\rho)<\mathcal{D}_\text{SSEP}(\rho)$; in contrast, the relaxation to the equilibrium proceeds faster for the RP since $D(\rho)> D_\text{SSEP}=1$. 

We emphasize that even for the one-dimensional SSEP, the normal diffusion behavior \eqref{var_tag} holds in finite systems, e.g. on the ring or in the case of open boundary conditions. In these settings, the self-diffusion constant has been computed (see \cite{spider1d} for one such calculation) and it was found to scale as $L^{-1}$ with system size, so it vanishes in the thermodynamic limit. Note also that the two growth laws, $\sqrt{t}$ and $L^{-1}t$, match at $t_c\sim L^2$ which is the usual diffusion time. The self-diffusion coefficient was also computed \cite{DEM93,DEM95,DM97} for the asymmetric SEP in finite systems where it was found to scale as $L^{-1/2}$ with system size, reflecting the $\langle X^2\rangle_c\sim t^{2/3}$ growth law of the variance for infinite systems.

\section{Lattice gases with an extra next-nearest-neighbor repulsion}
\label{NNN}

The RP studied in the previous sections was based on the zero-temperature dynamics compatible with Hamiltonian \eqref{Ham} accounting for the nearest-neighbor repulsion. In this section we modify the Hamiltonian to include the repulsion with next-nearest-neighbors, namely we consider the RP with Hamiltonian \eqref{Ham:2}. The coupling constants are assumed to obey $J_1>J_2>0$, so that both interactions are repulsive and the strength decays with separation. The precise values of the coupling constants are immaterial, the above assumptions suffice to devise a unique zero-temperature dynamics. The general property of zero-temperature dynamics is that an attempted hop to an empty site is performed if the energy is not raised and rejected otherwise. Thus the dynamical rules can be summarized as follows:
\begin{enumerate}
\item If an attempted hop would decrease the number of nearest-neighbor pairs of particles, it is always performed. 
\item If  the number of nearest-neighbor pairs remains the same, and the number of next-nearest-neighbors would not increase, the hop is always performed.
\item Otherwise, the attempted hop is never performed. 
\end{enumerate}

\subsection{Steady states and current}

The system wanders on a phase space of maximal-island configurations, all of them occurring with the same probability. Admissible maximal-island configurations are now different in the regions $0<\rho<\tfrac{1}{3}$, $~\tfrac{1}{3}<\rho<\tfrac{1}{2}$, $~\tfrac{1}{2}<\rho<\tfrac{2}{3}$, and $\tfrac{2}{3}<\rho< 1$. It suffices to examine two low-density regions:  $\rho<\tfrac{1}{3}$ and $\tfrac{1}{3}<\rho<\tfrac{1}{2}$.

When $\rho<\tfrac{1}{3}$, the steady states are maximal-island configurations with islands of vacant sites of length $\geq 2$:
\begin{equation*}
\bullet\,\circ\,\circ\,\bullet\,\circ\,\circ\,\circ\,\bullet\,\circ\,\circ\,\circ\,\circ\,\bullet\,\circ\,\circ\,\bullet\,\circ\,\circ\,\circ\,\,\circ
\end{equation*}
The total number of configurations of type 
\begin{equation}
\label{2}
\bullet\,\circ\,\circ\,\,\overbrace{\downarrow\underbrace{\circ\downarrow\circ\downarrow\circ\downarrow\circ\downarrow\circ\downarrow\circ\downarrow\circ}_{V-N-2 ~\text{vacancies}}\downarrow}^{N-1~ \text{doublets}}\,\,\circ
\end{equation}
($\downarrow$ denotes a possible location of the doublet $\bullet\,\circ$) is equal to $\binom{V-N-1}{N-1}$. The total number of admissible maximal-island configurations is therefore
\begin{equation}
\label{conf:2}
\mathcal{C} = \frac{L}{N} \binom{V-N-1}{N-1}\,, \quad \rho<\tfrac{1}{3}
\end{equation}
where the $\tfrac{L}{N}$ factor allows us to consider configurations \eqref{2} starting with the particle. To compute the current (in the asymmetric version) we merely count the number of configurations 
\begin{equation*}
\bullet\,\circ\,\circ\,\circ\,\,\overbrace{\downarrow\underbrace{\circ\downarrow\circ\downarrow\circ\downarrow\circ\downarrow\circ\downarrow\circ\downarrow\circ}_{V-N-3 ~\text{vacancies}}\downarrow}^{N-1~ \text{doublets}}\,\,\circ
\end{equation*}
and divide by the total number of configurations \eqref{conf:2}. We obtain
\begin{equation}
\label{current:2}
J = \frac{N}{L}\,\,\frac{\binom{V-N-2}{N-1}}{\binom{V-N-1}{N-1}} = \frac{N(V-2N)}{L(V-N-1)}
\end{equation}

When $\tfrac{1}{3}<\rho<\tfrac{1}{2}$, admissible maximal-island configurations have islands of vacant sites of length 1 or 2:
\begin{equation}
\label{region}
\bullet\,\circ\,\circ\,\bullet\,\circ\,\circ\,\bullet\,\circ\,\bullet\,\circ\,\circ\,\bullet\,\circ\,\circ\,\bullet\,\circ\,\bullet\,\circ\,\bullet\,\,\circ
\end{equation}
Counting the total number of configurations of type
\begin{equation}
\label{12}
\bullet\,\circ\,\,\overbrace{\downarrow\bullet\,\circ\downarrow\bullet\,\circ\downarrow\bullet\,\circ\downarrow\bullet\,\circ\downarrow\bullet\,\circ\downarrow\bullet\,\circ\downarrow\bullet\,\circ\downarrow}^{V-N~\text{vacancies}}
\end{equation}
yields $\binom{N}{V-N}$. [In \eqref{12}, we denote  by $\downarrow$ a possible location of a vacancy.] The total number of admissible maximal-island configurations is therefore
\begin{equation}
\label{conf:12}
\mathcal{C} = \frac{L}{N} \binom{N}{V-N}\,, \quad \tfrac{1}{3}<\rho<\tfrac{1}{2}
\end{equation}
in agreement with the general prediction \eqref{conf:small}. To compute the current we count the number of configurations 
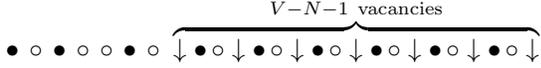
\begin{equation*}
\bullet\,\circ\,\bullet\,\circ\,\circ\,\bullet\,\circ\,\,\overbrace{\downarrow\bullet\,\circ\downarrow\bullet\,\circ\downarrow\bullet\,\circ\downarrow\bullet\,\circ\downarrow\bullet\,\circ\downarrow\bullet\,\circ\downarrow}^{V-N-1~\text{vacancies}}
\end{equation*}
and divide by \eqref{conf:12} to yield
\begin{equation}
\label{current:12}
J = \frac{N}{L}\,\,\frac{\binom{N-2}{V-N-1}}{\binom{N}{V-N}} = \frac{(2N-V)(V-N)}{L(N-1)}
\end{equation}

Taking the thermodynamic limit of \eqref{current:2} and \eqref{current:12}, extending to the high-density region, and collecting all the results we arrive at
\begin{equation}
\label{current:RP3}
J(\rho) = 
\begin{cases}
\frac{\rho(1-3\rho)}{1-2\rho}       & 0<\rho<\frac{1}{3}\\
\frac{(3\rho-1)(1-2\rho)}{\rho}     & \frac{1}{3}<\rho<\frac{1}{2}\\
\frac{(2\rho-1)(2-3\rho)}{1-\rho}  & \frac{1}{2}<\rho<\frac{2}{3}\\
\frac{(3\rho-2)(1-\rho)}{2\rho-1}  &\frac{2}{3}<\rho<1
\end{cases}
\end{equation}
This agrees with the announced general result \eqref{current:GRP} specialized to $m=2$.

\subsection{Correlation functions}

In the $\rho<\frac{1}{3}$ region, the pair correlation is easily computed when the separation does not exceed 2:
\begin{equation}
\label{nn:012}
\langle n_i^2\rangle=\rho, \quad  \langle n_i n_{i+1}\rangle = \langle n_i n_{i+2}\rangle = 0
\end{equation}

To compute $\langle n_i n_{i+\ell}\rangle$ when $\ell\geq 3$, we first consider configurations with $q-1$ particles and $\ell - q$ vacancies between sites $i$ and $i+\ell$:
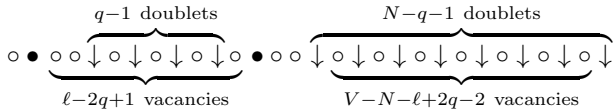
\begin{equation*}
\circ\bullet\,\underbrace{\circ\circ\overbrace{\downarrow\circ\downarrow\circ\downarrow\circ\downarrow}^{q-1~ \text{doublets}}\circ}_{\ell -2q+1~\text{vacancies}}\, \bullet\circ\circ\,\overbrace{\downarrow\underbrace{\circ\downarrow\circ\downarrow\circ\downarrow\circ\downarrow\circ\downarrow\circ\downarrow\circ}_{V-N-\ell+2q-2 ~\text{vacancies}}\downarrow}^{N-q-1~ \text{doublets}}
\end{equation*}
Particles at sites $i$ and $i+\ell$ are explicitly presented; other particles belong to $\bullet\,\circ$ doublets, possible locations of doublets are shown by $\downarrow$. The total number of such configurations is
\begin{equation}
\label{conf:q2}
\binom{\ell - 2q-1}{q-1}\binom{V-N-\ell+2q-1}{N-q-1}
\end{equation}
Summing over $q$ from $q=1$ to $q=\lfloor \ell/3\rfloor$ we get
\begin{equation*}
\langle n_i n_{i+\ell}\rangle = \frac{N}{L}\sum_{q=1}^{\lfloor \ell/3\rfloor}
\binom{\ell - 2q-1}{q-1}\,\frac{\binom{V-N-\ell+2q-1}{N-q-1}}{\binom{V-N-1}{N-1}}
\end{equation*}
which in the thermodynamic limit becomes
\begin{equation}
\label{2CF:q2}
\langle n_i n_{i+\ell}\rangle = \sum_{q=1}^{\lfloor \ell/3\rfloor} \binom{\ell - 2q-1}{q-1}\,
\frac{\rho^{q+1} (1-3\rho )^{\ell-3q}}{(1-2\rho)^{\ell-2q}}
\end{equation}

In the $\frac{1}{3}<\rho<\frac{1}{2}$ region analogous computations give
\begin{equation}
\label{2CF:q12}
\langle n_i n_{i+\ell}\rangle = \sum_{q=\lfloor \frac{\ell+2}{3}\rfloor}^{\lfloor \ell/2\rfloor} \binom{q}{\ell - 2q}\,
\frac{(1-2\rho)^{\ell-2q}}{\rho^{q-1}(1-3\rho )^{\ell-3q}}
\end{equation}

For higher-order correlation functions, Eqs.~\eqref{3CF}--\eqref{high_CF} continue to hold. This is valid in the entire density range. 

\subsection{Compressibility}

When $\rho<\frac{1}{3}$, we use the definition \eqref{compress:def} together with Eq.~\eqref{nn:012} and write the compressibility in the form
\begin{equation}
\label{chi:sum}
\chi=\rho(1-\rho)-4\rho^2+2\sum_{\ell\geq 3} \left[\langle n_i n_{i+\ell}\rangle - \rho^2\right]
\end{equation}
To compute the sum in \eqref{chi:sum} we generalize it to
\begin{equation}
\label{chi:sum2}
S(\rho,\lambda)=\sum_{\ell\geq 3} \left[\langle n_i n_{i+\ell}\rangle - \rho^2\right]\lambda^\ell
\end{equation}
and then deduce the necessary sum by taking the $\lambda\uparrow 1$ limit. The advantage of this trick is that we can split the sum \eqref{chi:sum2} into two sums which are both converging when $\lambda<1$. One gets (see Appendix~\ref{compress})
\begin{equation}
\label{chi:RP2}
\chi(\rho) = \rho(1-2\rho) (1-3\rho), \quad  \rho<\tfrac{1}{3}
\end{equation}

Similar calculations (see Appendix~\ref{compress}) give 
\begin{equation}
\label{chi:RP12}
\chi(\rho) = \rho(1-2\rho) (3\rho-1), \quad  \tfrac{1}{3}<\rho<\tfrac{1}{2}
\end{equation}

Using mirror symmetry we then determine $\chi(\rho)$ in the entire density range:
\begin{equation}
\label{chi:RP3}
\chi(\rho) = 
\begin{cases}
\rho(1-3\rho)(1-2\rho)        & 0<\rho<\frac{1}{3}\\
\rho(3\rho-1)(1-2\rho)        & \frac{1}{3}<\rho<\frac{1}{2}\\
(1-\rho)(2\rho-1)(2-3\rho)  & \frac{1}{2}<\rho<\frac{2}{3}\\
(1-\rho)(2\rho-1)(3\rho-2)  &\frac{2}{3}<\rho<1
\end{cases}
\end{equation}

\subsection{Diffusion and self-diffusion}

Using \eqref{EGK:simple} together with \eqref{current:RP3} and \eqref{chi:RP3} we compute the diffusion coefficient
\begin{equation}
\label{diff:RP3}
D(\rho) = 
\begin{cases}
(1-2\rho)^{-2}     & 0<\rho<\frac{1}{3}\\
\rho^{-2}             & \frac{1}{3}<\rho<\frac{1}{2}\\
(1-\rho)^{-2}       & \frac{1}{2}<\rho<\frac{2}{3}\\
(2\rho-1)^{-2}     &\frac{2}{3}<\rho<1
\end{cases}
\end{equation}
in agreement with the announced result \eqref{diff:GRP} specialized to $m=2$.

We can apply the Green-Kubo formula \eqref{EGK:simple} since the gradient condition holds; it does not hold manifestly (as for the SSEP), but it holds at equilibrium. This can be verified by straightforward analysis. 

The variance of the tagged particle is described by \eqref{var_tag_1d}. Combining the general relation 
 \eqref{var_tag_conj} with \eqref{chi:RP3} and \eqref{diff:RP3} we compute the amplitude in \eqref{var_tag_1d}:
\begin{equation}
\label{self-diff:RP3}
\mathcal{D}(\rho) = \frac{2}{\sqrt{\pi}\,\rho^2} \times
\begin{cases}
\rho(1-3\rho)              & 0<\rho<\frac{1}{3}\\
(3\rho-1)(1-2\rho)      & \frac{1}{3}<\rho<\frac{1}{2}\\
(2\rho-1)(2-3\rho)      & \frac{1}{2}<\rho<\frac{2}{3}\\
(1-\rho)(3\rho-2)        &\frac{2}{3}<\rho<1
\end{cases}
\end{equation}

\section{Generalized Repulsion Processes}
\label{GRP}

Consider now the generalized repulsion process (GRP) characterized by the Hamiltonian \eqref{Ham:GRP}, with coupling constants satisfying the set of constraints \eqref{JJJ}. One can verify the structure \eqref{islands} of admissible maximal-island configurations in regions \eqref{regions} by extending the arguments given for the RP. One can also check that configurations occur with equal probabilities; these probabilities depend on the region and on the density, and are equal to the inverse of the total number of configurations. The derivations of the announced results for the total number of configurations, Eqs.~\eqref{conf:tiny}--\eqref{conf:small}, and the steady state current \eqref{current:GRP} are also straightforward. The expression \eqref{diff:GRP} for the diffusion coefficient then follows from the Green-Kubo formula \eqref{EGK:simple} when we combine it with expressions \eqref{current:GRP} for the current and with formula 
\begin{equation}
\label{compress:GRP}
\chi = 
\begin{cases}
\rho [1-(m + 1)\rho][1- m \rho]& 0<\rho<\frac{1}{m+ 1}\\
\rho [(k + 1)\rho-1][1-k\rho] & \frac{1}{k+1}<\rho<\frac{1}{k}
\end{cases}
\end{equation}
giving the compressibility; $k=2,\ldots, m$ in Eq.~\eqref{compress:GRP}. To justify the applicability of the Green-Kubo formula \eqref{EGK:simple} we use again the gradient condition; checking of the validity of the gradient condition at equilibrium is tedious, but straightforward. 

To complete the derivation we must establish Eq.~\eqref{compress:GRP}. The approach that has been used previously is still applicable: One determines the connected pair correlation function and then one finds the compressibility via the definition \eqref{compress:def}. A long calculation was necessary even when $m=2$. To circumvent the necessity for such a calculation, we employ a different strategy. The compressibility can be found (see e.g. \cite{D07}) from relation
\begin{equation}
\label{comp_F}
\chi \,\frac{d^2 F}{d\rho^2} = 1
\end{equation}
Here $F$ is a free energy per lattice site:
\begin{equation}
\label{F_def}
F = - \lim_{L\to\infty} \frac{1}{L}\,\ln \mathcal{C}
\end{equation}
The computation of the total number of configurations $\mathcal{C}$ is much simpler than the computation of the connected pair correlation function, and then from Eqs.~\eqref{comp_F}--\eqref{F_def} we directly deduce the compressibility. 

For the RP, we get
\begin{equation*}
F=\rho\ln\rho-(1-\rho)\ln(1-\rho)+(1-2\rho)\ln(1-2\rho)   
\end{equation*}
in the low-temperature region. Using this result in conjunction with \eqref{comp_F} we re-derive \eqref{chi:RP} thereby providing a useful check of the consistency of our results [Eqs.~\eqref{nn:simple} and \eqref{nn2:high}] for the connected pair correlation function. 

Similarly for the RP involving next-nearest-neighbor repulsive interactions,  we use \eqref{conf:2} and \eqref{conf:12} to compute
\begin{equation*}
F=\rho\ln\rho - (1-2\rho)\ln(1-2\rho) + (1-3\rho)\ln(1-3\rho)
\end{equation*}
for $0<\rho<\tfrac{1}{3}$ and
\begin{equation*}
F=  (1-2\rho)\ln(1-2\rho) - \rho\ln\rho + (3\rho-1)\ln(3\rho-1)
\end{equation*}
for $\tfrac{1}{3}<\rho<\tfrac{1}{2}$. Using these results in conjunction with \eqref{comp_F} we re-derive \eqref{chi:RP3}. 

Generally, one can use Eqs.~\eqref{conf:tiny}--\eqref{conf:small} to establish the free energy. This leads to Eq.~\eqref{compress:GRP} for arbitrary $m$. 

\section{Discussion}
\label{Discussion}

We studied dynamical behaviors of one-dimensional stochastic lattice gases. We assumed that each site is occupied by at most one particle and that particles interact through repulsive forces whose span can be arbitrary. We endowed the system with a zero-temperature dynamics, so that the hops which would have led to the increase of energy are forbidden.  When the strength of interactions monotonically and rapidly decreases with the separation between the particles, namely the coupling constants satisfy the set of constraints \eqref{JJJ}, interactions can be treated in a lexicographic order.  In the case of symmetric nearest-neighbor hopping, we analytically determined the density-dependent diffusion coefficient. We also computed the variance of the displacement of a tagged particle. 

The repulsion process (RP) with nearest-neighbor interactions is related to a few other lattice gases. One such model is the facilitated exclusion process \cite{GKR} which is well-defined in the high-density regime and has the same structure of steady states and the same current as the RP, even though away from the steady state the dynamical rules of the two processes differ. Another is the KLS model \cite{KLS84} which reduces to the RP at a certain point of the parameter space. Our the results for the current can be extracted from the earlier work \cite{Krug01} on the KLS model, and they apparently were known much earlier \cite{B91}. The correlation functions, the density-dependent diffusion coefficient, and the amplitude characterizing self-diffusion haven't been investigated even in the context of the RP with nearest-neighbor interactions. Most of our results for the RP with nearest-neighbor interactions apparently admit a generalization to the KLS model. The tools are different, namely one employs a transfer-matrix technique \cite{Krug01,Bunin13}; the emerging results are very similar \cite{PK13}.  

The vast majority of lattice gases which have ever been studied involve only nearest-neighbor interactions. Therefore the generalized repulsion processes (GRP) with interactions whose span can be arbitrary form an interesting class of lattice gases with long-ranged interactions which, remarkably, admit the exact analyses. 

In this work we limited ourselves to the GRP on the ring. It would be interesting to study the GRP with asymmetric hopping in an open setting. Interesting behaviors have been reported for lattice gases with the current $J(\rho)$ having one or two maxima \cite{Krug91,Krug01}. For the GRP we have potentially many maxima, viz. $2m$ maxima for the system with Hamiltonian \eqref{Ham:GRP}, which might result in a rich phase diagram. 

Another avenue for future work is to look for large deviations \cite{D07,Bertini}. Interesting singular behaviors of the large deviation function have been recently found \cite{Bunin13} in the context of the KLS model, in the point in the parameter space close to the RP. Similar singular behaviors probably arise for the RP, and perhaps more subtle behaviors will appear for the GRPs. In Appendix~\ref{LD} we specialize 
the macroscopic fluctuation theory \cite{Bertini} describing large deviations in diffusive lattice gases to repulsion processes.  

Finally we note that it would be interesting to perform an analytical investigation of the driven GRP in the setting with open boundaries. The influence of repulsion can be striking, e.g. very rich phase diagrams were found in similar models (see \cite{Krug01,Mario} and references terein).  

\section*{Acknowledgments}

I am grateful to Uttam Bhat, Kirone Mallick, Baruch Meerson, and Sid Redner for discussions.

\appendix
\section{Higher-order connected correlation functions}
\label{higher}

Here we establish explicit expressions for the connected correlation functions $\langle 12345 \rangle_c$ and $\langle 123456 \rangle_c$. We then compute the number of exponential terms arising in the $p-$particle connected correlation function for arbitrary $p$. 

Let's begin with already known connected correlation functions given by Eqs.~\eqref{nn:simple},  \eqref{nnn:simple}, and  \eqref{nnnn}. It is useful to think about terms in these equations as representing contributions from the diagrams: $\widehat{12}$ in the case of \eqref{nn:simple}; $\widehat{123}$ in the case of \eqref{nnn:simple}; $\widehat{12}\,\widehat{34}$ and $\widehat{1234}$ in the case of \eqref{nnnn}. Continuing this line of reasoning we see that diagrams 
\begin{equation*}
\widehat{12}\,\, \widehat{345} \qquad \widehat{123}\,\, \widehat{45} \qquad  \widehat{12345}
\end{equation*}
contribute to $\langle 12345\rangle_c \equiv \langle n_{i_1} n_{i_2} n_{i_3} n_{i_4} n_{i_5} \rangle_c$. Therefore
\begin{eqnarray}
\label{n5}
\langle 12345 \rangle_c &=&A R^{i_2-i_1+i_5-i_3}+AR^{i_3-i_1+i_5-i_4} \nonumber\\
&+& BR^{i_5-i_1}, \quad         R =-\frac{\rho}{1-\rho}
\end{eqnarray}
The amplitudes in front of the terms in the top line of Eq.~\eqref{n5} are equal each other due to symmetry. The ratio $R =-\tfrac{\rho}{1-\rho}$ lies in the range $-1<R<0$ in the low-density region $\rho<\tfrac{1}{2}$. 

Similarly, the diagrams 
\begin{equation*}
\widehat{12}\,\, \widehat{34}\,\, \widehat{56}\,  \quad \widehat{123}\,\, \widehat{456}\, \quad  \widehat{12}\,\widehat{3456}\, \quad \widehat{1234}\,\, \widehat{56}\, \quad \widehat{123456} 
\end{equation*}
contribute to $\langle 123456\rangle_c$. This leads to
\begin{eqnarray}
\label{n6}
\langle 123456\rangle_c &=& UR^{i_6-i_5+i_4-i_3+i_2-i_1}
+V R^{i_6-i_4 + i_3-i_1}\nonumber\\
&+&V R^{i_6-i_3+i_2-i_1} + V R^{i_6-i_5+i_4-i_1}  \nonumber\\
&+&WR^{i_6-i_1}
\end{eqnarray}
The amplitudes appearing in the second line of Eq.~\eqref{n6} are equal due to symmetry, and they coincide with one of the amplitudes in the first line (we don't know a conceptual explanation for this equality). 

To determine the amplitudes $A$ and $B$ in Eq.~\eqref{n5} we set $1=2$ (that is, $i_1=i_2$). Using the definition \eqref{high_CF} together with relation
\begin{eqnarray}
\label{n2}
(n-\rho)^2 &=& n^2-2n\rho+\rho^2 = n-2n\rho+\rho^2 \nonumber\\
                 &=& (1-2\rho)(n-\rho)+\rho(1-\rho)
\end{eqnarray}
we find
\begin{eqnarray*}
\langle 22345 \rangle_c &=& (1-2\rho)\langle 2345 \rangle_c + \rho(1-\rho)\langle 345 \rangle_c\\
                                      &=& (1-2\rho)[\rho(1-\rho)]^2R^{i_3-i_2+i_5-i_4}\\
                                      &+&(1-2\rho)^3 \rho(1-\rho)R^{i_5-i_2}\\
                                      &+&(1-2\rho)[\rho(1-\rho)]^2R^{i_5-i_3}
\end{eqnarray*}
On the other hand, Eq.~\eqref{n5} reduces to 
\begin{equation*}
\langle 22345 \rangle_c = A R^{i_5-i_3}+AR^{i_3-i_2+i_5-i_4} + BR^{i_5-i_2}
\end{equation*}
when $i_1=i_2$. Comparing these two results we get 
\begin{equation*}
A = (1-2\rho)[\rho(1-\rho)]^2, \quad 
B = (1-2\rho)^3 \rho(1-\rho)
\end{equation*}

Similarly we set  $1=2$ in Eq.~\eqref{n6} and reduce it to
\begin{eqnarray}
\label{6:2}
\langle 223456\rangle_c &=& UR^{i_6-i_5+i_4-i_3}+ V R^{i_6-i_4 + i_3-i_2}\nonumber\\
&+&V R^{i_6-i_3} + V R^{i_6-i_5+i_4-i_2} \nonumber\\
&+&WR^{i_6-i_2}
\end{eqnarray}
On the other hand, using \eqref{high_CF} and \eqref{n2} we obtain
\begin{eqnarray*}
\langle 223456 \rangle_c &=& (1-2\rho)\langle 23456 \rangle_c + \rho(1-\rho)\langle 3456 \rangle_c\\
                                      &=& (1-2\rho)A\left(R^{i_6-i_4+i_3-i_2}+R^{i_6-i_5+i_4-i_2}\right)\\
                                      &+&(1-2\rho)BR^{i_6-i_2}+[\rho(1-\rho)]^3 R^{i_6-i_5+i_4-i_3}\\
                                      &+&(1-2\rho)^2[\rho(1-\rho)]^2R^{i_6-i_3}
\end{eqnarray*}
Comparing this with \eqref{6:2} we determine the amplitudes 
\begin{equation*}
\label{ampl}
\begin{split}
U &= [\rho(1-\rho)]^3\\
V &= (1-2\rho)^2[\rho(1-\rho)]^2\\
W &= (1-2\rho)^4 \rho(1-\rho)
\end{split}
\end{equation*}

Consider now the general case and denote by $D_p$ the number of diagrams (equivalently, the number of exponential terms) appearing in the $p-$particle connected correlation function. The number of diagrams starting with $\widehat{12}$ is equal to $D_{p-2}$, the number of diagrams starting with $\widehat{123}$ is equal to $D_{p-3}$, etc. Therefore
\begin{equation*}
D_p = D_{p-2}+D_{p-3}+\ldots+D_2+1
\end{equation*}
which can be re-written as the recurrence 
\begin{equation}
D_p = D_{p-2}+D_{p-1}
\end{equation}
defining the Fibonacci numbers. The standard initial condition for the Fibonacci numbers is $F_1=F_2=1$, while in our case $D_2=D_3=1$. Hence $D_p=F_{p-1}$. 

\section{Repulsion processes and gradient condition}
\label{gradient}

Zero-temperature dynamics which are compatible with Hamiltonian \eqref{Ham} and involve only nearest-neighbor hopping proceed according to the following rules:
\begin{enumerate}
\item If a hop (to a neighboring empty site) would result in decrease of the number of nearest-neighbor pairs of particles, $\Delta H = -1$, it is always performed. 
\item If a hop would cause neither decrease nor increase of the number of nearest-neighbor pairs of particles, $\Delta H = 0$, it is performed with probability $p$. 
\item If a hop would increase of the number of nearest-neighbor pairs of particles, $\Delta H = 1$, it is never performed.
\end{enumerate}

The probability $p$ is a parameter of the model. Three values of $p$ play a special role:
\begin{itemize}
\item $p=0$.  Only strictly energy lowering hops are allowed. In this case, the system quickly gets trapped in a jammed configuration. 
\item $p=\frac{1}{2}$.  In zero-temperature spin-flip dynamics, this special value is prescribed by Glauber dynamics, and it is known to simplify analysis in a few tractable cases \cite{book}. 
\item $p=1$. This value is most efficient in simulations, and it is typically used in Metropolis and heat-bath algorithms. 
\end{itemize}

As long as $p>0$, its value usually plays a minor role. In this paper we have chosen $p=1$ although, similarly to zero-temperature spin-flip dynamics, the choice $p=\frac{1}{2}$ is theoretically advantageous.  To see why this is so, let us consider the current $J_{i,i+1}$ through the bond $(i,i+1)$. This current is fully determined by occupation variables $n_{i-1},\,n_{i},\,n_{i+1},\,n_{i+2}$. When $p=\frac{1}{2}$, the current reads (we assume that the hopping is symmetric):
\begin{equation}
\label{Glauber}
2J_{i,i+1}=n_i-n_{i+1}+(n_{i-1}-n_{i+2})(n_i-n_{i+1})^2
\end{equation}
while for $p=1$ one gets
\begin{eqnarray}
\label{Metropolis}
J_{i,i+1} &=& n_{i-1}n_i(1-n_{i+1})-(1-n_i)n_{i+1}n_{i+2}\nonumber\\
              &+&  (n_i-n_{i+1})(1-n_{i-1})(1-n_{i+2})
\end{eqnarray}

The current corresponding to the $p=\frac{1}{2}$ case is not merely expressed by a more compact formula than the current corresponding to $p=1$. A crucial property is that for $p=\frac{1}{2}$ the current has a gradient form. Indeed, massaging \eqref{Glauber} we re-write it as
\begin{eqnarray*}
2J_{i,i+1} &=& [n_i-n_{i+1}] + [n_{i-1}n_{i+1}-n_in_{i+2}]\\
&+& [n_{i-1}n_i-n_in_{i+1}]+[n_in_{i+1}-n_{i+1}n_{i+2}]\\
&-&2[n_{i-1}n_in_{i+1}-n_i n_{i+1} n_{i+2}]
\end{eqnarray*}
where each term in the square brackets is a (discrete) gradient of some local function of ${\bf n}=\{n_j\}$. 
For $p=1$, and for any $p\ne \frac{1}{2}$, the current does not have a gradient form. For instance, for $p=1$ we can re-write \eqref{Metropolis} as
\begin{eqnarray*}
J_{i,i+1} &=& [n_i-n_{i+1}] + [n_{i-1}n_{i+1}-n_in_{i+2}]\\
&-& [n_{i-1}n_in_{i+1}-n_i n_{i+1} n_{i+2}]\\
&+&n_{i-1}n_i n_{i+2}-n_{i-1} n_{i+1} n_{i+2}
\end{eqnarray*}
Each term in  the square brackets is a gradient; the terms in the last line cannot be converted into a gradient form. 

If the current across the bond admits a gradient representation, the computation of the diffusion coefficient greatly simplifies, namely the integral term in Eq.~\eqref{EGK} vanishes \cite{Spohn91}, and we can a simple formula \eqref{EGK:simple}. This formally justifies the results of Sect.~\ref{DSD} for the RP with $p=\frac{1}{2}$. We now recall that the hydrodynamic regime is very close to the (local) equilibrium where energy lowering hops are no longer possible. Therefore there is no distinction between the dynamics with any $p>0$, all hops occur with the same rate $p$. Therefore the expression for the diffusion coefficient is actually the same (apart from the trivial rescaling, $D\to pD$).

\section{Compressibility}
\label{compress}

Let us split the sum \eqref{chi:sum2} into two sums 
\begin{equation*}
S(\rho,\lambda)= \sum_{\ell\geq 3} \langle n_i n_{i+\ell}\rangle\, \lambda^\ell   - \rho^2\sum_{\ell\geq 3} \lambda^\ell = S_1 - S_2
\end{equation*}
To compute $S_1$ we use \eqref{2CF:q2}, write $\ell=3q+m$, change the order of summation, and after a bit of massaging we get
\begin{eqnarray*}
S_1 &=& \rho \sum_{q\geq 1}\left(\frac{\rho \lambda^3}{1-2\rho}\right)^q \sum_{m\geq 0} \binom{q-1+m}{q-1}\,
\left[\lambda\, \frac{1-3\rho}{1-2\rho}\right]^m\\
&=&\rho \sum_{q\geq 1}\left(\frac{\rho \lambda^3}{1-2\rho}\right)^q \left[1-\lambda\, \frac{1-3\rho}{1-2\rho}\right]^{-q}\\
&=&\frac{\rho^2\lambda^3}{1-2\rho-\lambda(1-3\rho)-\rho\lambda^3}
\end{eqnarray*}
Therefore
\begin{equation*}
S(\rho,\lambda)=\frac{\rho^2\lambda^3}{1-2\rho-\lambda(1-3\rho)-\rho\lambda^3}-\frac{\rho^2\lambda^3}{1-\lambda}
\end{equation*}
from which we deduce $\lim_{\lambda\to 1}S(\rho,\lambda)=3\rho^3$. Plugging this into \eqref{chi:sum} we arrive at the announced result \eqref{chi:RP2}. 

Similarly one can determine $\chi$ in the $\frac{1}{3}<\rho<\frac{1}{2}$ region. As previously, it suffices to compute
\begin{equation}
\label{chi:lambda}
\rho(1-\rho)+2S_1(\rho,\lambda) - \frac{2\rho^2\lambda}{1-\lambda}
\end{equation}
where $S_1=\sum_{\ell\geq 2} \langle n_i n_{i+\ell}\rangle\, \lambda^\ell$.  We plug  \eqref{2CF:q12} into $S_1$, write $\ell=2q+m$, and change the order of summation to yield
\begin{eqnarray*}
S_1 &=& \sum_{q\geq 1}\frac{\lambda^{2q}}{\rho^{q-1}} \sum_{m=0}^q \binom{q}{m}\,
[\lambda(1-2\rho)]^m (3\rho-1)^{q-m}\\
&=&\sum_{q\geq 1}\frac{\lambda^{2q}}{\rho^{q-1}}[\lambda(1-2\rho)+3\rho-1]^m\\
&=&-\rho+\frac{\rho^2}{\rho-\lambda^2[\lambda(1-2\rho)+3\rho-1]}
\end{eqnarray*}
Substituting this result into \eqref{chi:lambda} and taking the $\lambda\uparrow 1$ limit we arrive at the announced result \eqref{chi:RP12}. 

\section{Large deviations}
\label{LD}

The macroscopic fluctuation theory (MFT) has been recently developed to probe large deviations in lattice gases \cite{Bertini}.  Mathematically, one must solve two coupled partial differential equations
\begin{equation}
\begin{split}
\partial_t \rho &= \partial_x \left[D(\rho)\, \partial_x \rho\right]
-  \partial_x \left[\sigma(\rho)\, \partial_x p\right]\\
\partial_t p & = - D(\rho) \partial_{xx} p
- \tfrac{1}{2}\sigma^{\prime}(\rho)\!\left(\partial_x p\right)^2
\end{split}
\end{equation}
for the density field $\rho(x,t)$  and the conjugate momentum field $p(x,t)$. These equations contain the diffusion coefficient $D(\rho)$ and the quantity $\sigma(\rho)$ which characterizes the variance of the flux. More precisely, consider a one-dimensional system of a very large length $L$ with open boundaries which are in contact with reservoirs of particles with density $\rho$. The variance of the the integrated flux gives $\sigma(\rho)$:
\begin{equation}
\label{sigma_def}
\lim_{t\to\infty} \frac{\langle I^2\rangle}{t} = \frac{\sigma(\rho)}{L}
\end{equation}
The equilibrium origin of the quantities $D(\rho)$ and $\sigma(\rho)$ is emphasized by relation 
\begin{equation}
\label{FDT}
\frac{d^2 F(\rho)}{d\rho^2} = \frac{2D(\rho)}{\sigma(\rho)}
\end{equation}
which follows \cite{Spohn91,D07} from the fluctuation-dissipation theorem. Combining \eqref{FDT} with \eqref{comp_F} we arrive at  $\sigma = 2D\chi$, which simplifies to 
\begin{equation}
\sigma(\rho) = 2J(\rho)
\end{equation}
for all GRP. Interestingly, $\sigma(\rho)$ describes the symmetric version of the process, while $J(\rho)$ is the steady state current in the totally asymmetric version. We have computed $J(\rho)$, see \eqref{current:RP} and \eqref{current:GRP}, and hence we know $\sigma(\rho)$ for an infinite class of repulsion processes.

\end{document}